\documentclass[aps,prd,twocolumn,groupedaddress,showpacs]{revtex4-1}
\usepackage{epsf,epsfig,graphicx}
\usepackage{amsmath}
\usepackage{amssymb}
\usepackage{tikz}

\def\edth{\;\raise1.0pt\hbox{$'$}\hskip-6pt\partial\;}
\def\baredth{\;\overline{\raise1.0pt\hbox{$'$}\hskip-6pt
\partial}\;}
\def\gsim{~\rlap{$>$}{\lower 1.0ex\hbox{$\sim$}}}

\newcommand{\be}{\begin{equation}}
\newcommand{\ee}{\end{equation}}
\newcommand{\bw}{\begin{widetext}}
\newcommand{\ew}{\end{widetext}}

\newcommand{\intinf}{\int_{-\infty}^{\infty}}
\newcommand{\suml}{\sum_{l=0}^{\infty}}
\newcommand{\summ}{\sum_{m=-\ell}^{\ell}}

\usepackage{xcolor}

\AtBeginDocument{\renewcommand{\d}{\mathbf{d}}}

\begin{document}

\title{Timing-residual power spectrum of a polarized stochastic gravitational-wave background in pulsar-timing-array observation}

\author{Guo-Chin Liu$^{1}$ and Kin-Wang Ng$^{2,3}$}

\affiliation{
$^1$Department of Physics, Tamkang University, Tamsui, New Taipei City 25137, Taiwan\\
$^2$Institute of Physics, Academia Sinica, Taipei 11529, Taiwan\\
$^3$Institute of Astronomy and Astrophysics, Academia Sinica, Taipei 11529, Taiwan
}

\vspace*{0.6 cm}
\date{\today}
\vspace*{1.2 cm}

\begin{abstract}
We study the observation of stochastic gravitational-wave background (SGWB) made by pulsar-timing arrays in the spherical harmonic space. Instead of using the Shapiro time delay, we keep the Sachs-Wolfe line-of-sight integral for the timing residual of an observed pulsar. We derive the power spectrum of the timing residual, from which the overlap reduction functions and the bipolar spherical harmonics coefficients are constructed for the SGWB intensity and polarization anisotropies. We have reproduced the previous results, noting that we have developed a fast algorithm for computing accurate overlap reduction functions and the bipolar spherical harmonics coefficients for the linear-polarization anisotropy are worked out for the first time.
Our harmonic-space method is useful for future pulsar-timing-array observation on a few thousand pulsars and provides optimal estimators for testing the statistical isotropy of the SGWB.
\end{abstract}

\maketitle

\section{Introduction}

The detection of gravitational waves (GWs) emitted from the coalescence of binary black holes by the LIGO-Virgo experiment opens up a new era of GW astronomy and cosmology~\cite{ligo,ligo2019}. LIGO/Virgo is a ground-based laser interferometer that has two perpendicular detector-arms to measure GW strain amplitudes. Upcoming and future GW interferometry experiments include KAGRA, GEO600, Einstein Telescope, Cosmic Explorer, as well as space missions such as LISA, DECIGO, Taiji, and TianQin~\cite{ligo2050}, aiming to measure GWs at frequencies ranging from kilohertz to millihertz. Pulsar timing is another method to detect GWs by monitoring the arrival times of radio pulses from pulsars with ground-based radio telescopes~\cite{pta2019}. The line of sight from the telescope to the precise pulsar clock acts like a detector-arm sensitive to the passage of nanohertz GWs through the space. Current pulsar timing array (PTA) experiments, monitoring roughly 100 Galactic millisecond pulsars, include EPTA~\cite{EPTA},  NANOGrav~\cite{NANO}, and PPTA~\cite{PPTA}. The future SKA project will observe about 6000 Galactic millisecond pulsars to reach a sensitivity three to four orders of magnitude better than the current PTAs~\cite{SKA,SKA2}.

Stochastic gravitational wave background (SGWB) is a key science goal in GW experiments. GWs are very weakly interacting. The observation of SGWB enable us to see directly the physical processes that produce GWs in the early Universe such as distant compact binary coalescences, early-time phase transitions, cosmic string or defect networks, second-order primordial scalar perturbations, and inflationary GWs~\cite{romano}. The SGWB is predicted to be highly isotropic; however, it has been proposed that it could be anisotropic and even circularly or linearly polarized~\cite{alexander,satoh,sorbo,crowder2013,cusin,bartolo,pitrou,liu21,cai21}.

In search of SGWB in an interferometry network, the responses of a pair of detectors to the GW strain amplitude are correlated so as to filter out detector noises and increase the signal-to-noise ratio~\cite{romano}. In PTA observation, quadrupolar spatial correlations between pulsar pairs are used to identify the presence of a SGWB~\cite{pta2019}. 
Recently, the NANOGrav Collaboration~\cite{nanograv} has found strong evidence of a stochastic common-spectrum process across 45 ms pulsars, hinting at a SGWB with the spectral energy density of $\Omega_{\rm GW}\simeq 5.0\times 10^{-9}$ at a reference frequency of $32\,{\rm nHz}$. 
A consistent common-spectrum process has also been found in the second data release of the Parkes Pulsar Timing Array (PPTA)~\cite{PPTA21},
the Data Release 2 of the European Pulsar Timing Array (EPTA) covering a timespan up to 24 years~\cite{EPTA21}, 
and the second data release of the International Pulsar Timing Array (IPTA) synthesizing decadal-length pulsar-timing campaigns~\cite{IPTA22}.
However, the observations have not found statistically significant evidence that this process has quadrupolar interpulsar correlations.
Lately, using data from Advanced LIGO's and Advanced Virgo's third observing run (O3) combined with the earlier O1 and O2 runs, upper limits have been derived on an isotropic SGWB,  $\Omega_{\rm GW}<3.9\times 10^{-10}$ at $25\,{\rm Hz}$~\cite{ligo2101}, and on anisotropic SGWB, $\Omega_{\rm GW}< (0.56-9.7)\times 10^{-9} {\rm sr}^{-1}$~\cite{ligo2103}.

In this paper, we will study the observation of SGWB intensity and polarization anisotropies in PTAs. There has been a lot of theoretical studies on the pulsar-timing observation of SGWB~\cite{pta2019}.
Previous works have been mostly based on spatial correlation functions of the timing residuals from the Earth-term contribution for the intensity anisotropy~\cite{anholm,mingar13,gair14} and for the circular-polarization anisotropy~\cite{kato16}.
The modifications by the pulsar term to the intensity correlation functions have been discussed~\cite{mingar14}. 
In Ref.~\cite{qin19}, angular power spectra of the line-of-sight integral for the timing residual for the SGWB intensity and circular-polarization anisotropies have been derived in the total-angular-momentum formalism that uses chiral spherical gravitational waves.
Adopting the technique to expand the polarization basis tensors in terms of spin-weighted spherical harmonics~\cite{chu21}, 
a numerical scheme has been developed to calculate the correlation functions, 
being extended to including the linear-polarization anisotropy~\cite{chu2107}.
Here, following Ref.~\cite{ng21}, we will formulate the problem in the spherical harmonic space, directly deriving the power spectrum of the timing residual of an observed pulsar.
We will use the Sachs-Wolfe line-of-sight integral for the pulsar timing residual, adopted in Refs.~\cite{ng21,chu2107}, which have shown that the integral conveniently incorporates the effects of the pulsar term. 

Many papers have laid out schemes for measuring the SGWB intensity and polarization anisotropies 
in pulsar timing data analyses, such as the Bayesian parameter-estimation pipeline~\cite{taylor1,taylor2}, 
the spherical-harmonic power spectrum estimators~\cite{gair14,kam1,kam2}, 
and the Fisher matrix of observed pulsar pairs~\cite{smith1,smith2}.
Some of the limitations of using the spherical-harmonics approach such as being computationally demanding 
in the analysis pipeline and an inhomogeneous sky coverage of pulsars in real PTA observation 
have been discussed~\cite{smith1,smith2}. 
However, the results in this work will be useful in future PTA observation on a large number of pulsars.

The paper is organized as follows. We will introduce a polarized SGWB and its Stokes parameters in the next section, followed by a brief account of the pulsar timing in Sec.~\ref{pulsartiming}. In Secs.~\ref{spectrum} and \ref{spectrum2}, the power spectrum of the timing residual will be derived. We will obtain the overlap reduction functions in the celestial coordinates in Sec.~\ref{celestial} and in the computational frame in Sec.~\ref{computational}. We will briefly mention the bipolar spherical harmonics coefficients in Sec.~\ref{biposh}. Sec.~\ref{conclusion} is our conclusion.

\section{Polarized SGWB}

In the Minkowskian spacetime $(t,\vec{x})$, the metric perturbation $h_{ij}$ in the transverse traceless gauge depicts travelling GWs at the speed of light $c=\omega/k$. It can be expanded by Fourier modes as
\be
\label{eq:planwave}
h_{ij}(t,\vec{x}) = \sum_{A}\intinf \d f \int_{S^2} \d\hat{k} \;
    h_A(f,\hat{k}) \mathbf{e}^A_{ij}(\hat{k})
    e^{-2 \pi i f (t - \hat{k}\cdot \vec{x}/c)}\,,
\ee
where $A$ stands for the polarization of GWs with basis tensors $\mathbf{e}^A_{ij}(\hat{k})$, which are transverse to the propagation direction, $\hat{k}$. Here $h_{ij}$ is treated as real, so the Fourier components with negative frequencies are given by $h_A(-f,\hat{k})=h_A^*(f,\hat{k})$ for all $f\ge 0$. We define a SGWB as a collection of GWs satisfying the condition that $h_{ij}$ are random Gaussian fields with a statistical behavior completely characterized by
the two-point correlation function $\langle h_{ij}(t,\vec{x}_1) h_{ij}(t,\vec{x}_2) \rangle$, where the angle brackets denote
their ensemble averages. The ensemble averages of the Fourier modes have the following form				
\be
\label{eq:paa}
\langle h_{A}(f,\hat{k}) h^*_{A'}(f',\hat{k}') \rangle
= \delta(f-f') \delta(\hat{k}-\hat{k}')P_{AA'}(f,\hat{k}) \,,
\ee
where the spatial translational invariance dictates the delta function of their 3-momenta, $\delta(\vec{k}-\vec{k}')$. Note that the power spectra $P_{AA'}(f,\hat{k})$ remain to be direction dependent.

For GWs coming from the sky direction $-\hat{k}$ with wave vector $\vec{k}$, it is customary to write the polarization basis tensors in terms of the basis vectors in the spherical coordinates,
\begin{align}
\label{basisvector}
    \mathbf{e}^{+}(\hat{k}) &= \hat{\mathbf{e}}_\theta \otimes \hat{\mathbf{e}}_\theta  
                               - \hat{\mathbf{e}}_\phi \otimes \hat{\mathbf{e}}_\phi \,, \nonumber \\
    \mathbf{e}^{\times}(\hat{k}) &= \hat{\mathbf{e}}_\theta \otimes \hat{\mathbf{e}}_\phi  
                                    + \hat{\mathbf{e}}_\phi \otimes \hat{\mathbf{e}}_\theta  \,,
\end{align}
in which $\hat{\mathbf{e}}_\theta$, $\hat{\mathbf{e}}_\phi$, and $\hat{k}$ form a right-handed orthonormal basis.
Also, we can define the complex circular polarization basis tensors as
\begin{align}
    \mathbf{e}_{R} &= \frac{(\mathbf{e}_{+} + i \mathbf{e}_{\times})}{\sqrt{2}} \,,
   &\mathbf{e}_{L} &= \frac{(\mathbf{e}_{+} - i \mathbf{e}_{\times})}{\sqrt{2}} \,,
\end{align}
where $\mathbf{e}_{R}$ stands for the right-handed GW with a positive helicity while $\mathbf{e}_{L}$ stands for the left-handed GW with a negative helicity. The corresponding amplitudes in Eq.~(\ref{eq:planwave}) in the two different bases are related to each other via
\begin{align}
    h_{R} &= \frac{(h_{+} - i h_{\times})}{\sqrt{2}} \,,
   &h_{L} &= \frac{(h_{+} + i h_{\times})}{\sqrt{2}} \,.
\end{align}

Analogous to the case in electromagnetic waves~\cite{book:BornAndWolf}, the coherency matrix $P_{AA'}$ in Eq.~(\ref{eq:paa}) is related to the Stokes parameters, $I$, $Q$, $U$, and $V$ as 
\begin{align}
    I &= \left[ \langle h_R h_R^* \rangle + \langle h_L h_L^* \rangle \right] / 2 \,,\nonumber\\
    Q + iU &=  \langle h_L h_R^*  \rangle \,,\nonumber\\
    Q - iU &=  \langle h_R h_L^*  \rangle \,,\nonumber\\
    V &= \left[ \langle h_R h_R^* \rangle - \langle h_L h_L^* \rangle \right] / 2 \,\label{eq:spIQUV},
\end{align}
which are functions of the frequency $f$ and the propagation direction $\hat{k}$.
$I$ is the intensity, $Q$ and $U$ represent the linear polarization, and $V$ is the circular polarization.

\section{Pulsar timing}
\label{pulsartiming}

In the pulsar-timing observation, radio pulses from an array of roughly 100 Galactic millisecond pulsars are being monitored with ground-based radio telescopes. The redshift fluctuation of a pulsar in the pointing direction 
$\hat{e}$ on the sky is given by the Sachs-Wolfe effect~\cite{sachs},
\begin{equation}
z(\hat{e})= - {1\over 2}\int_{\eta_e}^{\eta_r}\d\eta\, \hat{e}^i \hat{e}^j \frac{\partial}{\partial\eta}h_{ij}(\eta, \vec x)\,,
\label{rsfze}
\end{equation}
where the lower (upper) limit of integration in the line-of-sight integral represents the point of emission (reception) 
of the radio pulse. The physical distance of the pulsar from the Earth is 
\be
D=c(\eta_r-\eta_e)\,, 
\ee
which is of order $1\,{\rm kpc}$.

The quantity that is actually observed in the pulsar-timing observation is the timing residual counted as
\begin{equation}
r(t)=\int_0^t \d t'z(t')\,,
\label{tresidual}
\end{equation}
where $t'$ denotes the laboratory time and $t$ is the duration of the observation. 
Using the laboratory time $t'$, we rewrite Eq.~(\ref{rsfze}) as
\begin{equation}
z(t',\hat{e})= - {1\over 2}\int_{t'+\eta_e}^{t'+\eta_r}\d\eta\, d^{ij}\frac{\partial}{\partial\eta}h_{ij}(\eta, \vec x)\,,
\label{labz}
\end{equation}
where the detector tensor is
\be
d^{ij}=\hat{e}^i \hat{e}^j\,.
\ee

\section{Timing-residual power spectrum}
\label{spectrum}

Then, replacing $\vec{x}$ by $c(\eta_r-\eta)\hat{e}$ in Eq.~(\ref{labz}) and using the spherical wave expansion~(\ref{eq:swexpansion}) 
for the phane wave~(\ref{eq:planwave}), Eq.~(\ref{tresidual}) becomes
\bw
\be
r(t,\hat{e})=2\pi\sum_{A}\intinf \d f \int_{S^2} \d\hat{k} \int_{\eta_e}^{\eta_r}\d\eta\,(1-e^{-2\pi i f t}) 
    h_A(f,\hat{k}) d^{ij} \mathbf{e}^A_{ij}(\hat{k})
    e^{-2 \pi i f \eta} \sum_{LM} i^L j_L[2\pi f (\eta_r-\eta)] Y_{LM}^*(\hat{k}) Y_{LM}(\hat{e})\,.
\ee
\ew
We expand 
\be
r(t,\hat{e})= \sum_{\ell m}  a_{\ell m} Y_{\ell m}(\hat{e})\,.
\ee
Defining $x=2\pi f (\eta_r-\eta)$, we have
\bw
\begin{align}
 a_{\ell m}&= \intinf \frac{\d f}{2\pi f}\,(1-e^{-2\pi i f t})  e^{-2 \pi i f \eta_r} 
 \sum_{A} \int_{S^2} \d \hat{k}\,h_A(f,\hat{k}) J^A_{\ell m}(fD,\hat{k})\,, 
\nonumber \\
J^A_{\ell m}(fD,\hat{k})&\equiv \sum_{LM} 2\pi\, i^{L} Y_{LM}^*(\hat{k}) \int_0^{2\pi fD/c}\d x\,e^{ix} j_{L}(x)
       \int_{S^2} \d \hat{e}\,d^{ij} \mathbf{e}^A_{ij}(\hat{k}) Y_{LM}(\hat{e}) Y^*_{\ell m}(\hat{e})\,.
\label{JAlm}       
\end{align}
\ew
The timing-residual correlation between a pair of Galactic pulsars $a$ and $b$ is constructed as
\bw
\be
\langle r(t_a,\hat{e}_a)r(t_b,\hat{e}_b) \rangle 
=\int_0^{t_a}\d t' \int_0^{t_b}\d t{''} \langle z(t',\hat{e}_a) z(t{''},\hat{e}_b)\rangle
=\sum_{\ell_1 m_1 \ell_2 m_2}
\langle a_{\ell_1 m_1} a^*_{\ell_2 m_2} \rangle Y_{\ell_1 m_1}(\hat{e}_a)  Y^*_{\ell_2 m_2}(\hat{e}_b)\,,
\label{EnAv}
\ee
where the ensemble average is given by Eq.~(\ref{eq:paa}) as
\be
\langle a_{\ell_1 m_1} a^*_{\ell_2 m_2} \rangle = \intinf \frac{\d f}{(2\pi f)^2}\,(1-e^{-2\pi i f t_a}) (1-e^{2\pi i f t_b}) 
\sum_{A_1 A_2} \int_{S^2} \d \hat{k}\, P_{A_1 A_2}(f,\hat{k}) J^{A_1}_{\ell_1 m_1}(fD_a,\hat{k}) J^{A_2 *}_{\ell_2 m_2}(fD_b,\hat{k})\,.
\ee
In terms of the Stokes parameters in Eq.~(\ref{eq:spIQUV}) and the definitions,
\begin{align}   
\mathbb{J}^{I}_{\ell_1 m_1 \ell_2 m_2} (fD_a,fD_b,\hat{k})&\equiv  J^{R}_{\ell_1 m_1}(fD_a,\hat{k}) J^{R *}_{\ell_2 m_2}(fD_b,\hat{k}) + J^{L}_{\ell_1 m_1}(fD_a,\hat{k}) J^{L *}_{\ell_2 m_2}(fD_b,\hat{k}) \,, \nonumber \\
\mathbb{J}^{V}_{\ell_1 m_1 \ell_2 m_2} (fD_a,fD_b,\hat{k})&\equiv  J^{R}_{\ell_1 m_1}(fD_a,\hat{k}) J^{R *}_{\ell_2 m_2}(fD_b,\hat{k}) - J^{L}_{\ell_1 m_1}(fD_a,\hat{k}) J^{L *}_{\ell_2 m_2}(fD_b,\hat{k}) \,, \nonumber \\ 
\mathbb{J}^{Q+iU}_{\ell_1 m_1 \ell_2 m_2} (fD_a,fD_b,\hat{k})&\equiv  J^{L}_{\ell_1 m_1}(fD_a,\hat{k}) J^{R *}_{\ell_2 m_2}(fD_b,\hat{k}) \,, \nonumber \\
\mathbb{J}^{Q-iU}_{\ell_1 m_1 \ell_2 m_2} (fD_a,fD_b,\hat{k})&\equiv  J^{R}_{\ell_1 m_1}(fD_a,\hat{k}) J^{L *}_{\ell_2 m_2}(fD_b,\hat{k}) \,,
\label{Jdefination}
\end{align}
we have
\be
\langle a_{\ell_1 m_1} a^*_{\ell_2 m_2} \rangle = \intinf \frac{\d f}{(2\pi f)^2}\,(1-e^{-2\pi i f t_a}) (1-e^{2\pi i f t_b}) 
\sum_{X=\{I,V,Q\pm iU\}} \int_{S^2} \d \hat{k}\, X(f,\hat{k})\, \mathbb{J}^{X}_{\ell_1 m_1 \ell_2 m_2} (fD_a,fD_b,\hat{k})\,.
\ee
\ew
We further expand the Stokes parameters in terms of ordinary and spin-weighted spherical harmonics as
\begin{align}
    I(f,\hat{k}) &= \sum_{\ell m}I_{\ell m}(f) \; Y_{\ell m}(\hat{k}) \,,\nonumber \\
    V(f,\hat{k}) &= \sum_{\ell m}V_{\ell m}(f) \; Y_{\ell m}(\hat{k}) \,,\nonumber \\
     (Q+iU)(f,\hat{k}) &= \sum_{\ell m}(Q+iU)_{\ell m}(f) \; _{+4}Y_{\ell m}(\hat{k}) \,,\nonumber \\
    (Q-iU)(f,\hat{k}) &= \sum_{\ell m}(Q-iU)_{\ell m}(f) \; _{-4}Y_{\ell m}(\hat{k}) \,,
\end{align}
where the specific combinations, $Q\pm iU$, make them become spin $\pm4$ objects so that we can expand them nicely by the corresponding spin-weighted spherical harmonics. A brief introduction to the spin-weighted spherical harmonics is found in Appendix~\ref{sec:spinweight}. 

Hence, we can express the timing-residual correlation in the following form
\bw
\be
\langle r(t_a,\hat{e}_a)r(t_b,\hat{e}_b) \rangle=
  \intinf \frac{\d f}{(2\pi f)^2}\,(1-e^{-2\pi i f t_a}) (1-e^{2\pi i f t_b}) 
   \sum_{X=\{I,V,Q\pm iU\}}
   \sum_{\ell m}  X_{\ell m}(f) \gamma_{\ell m}^{X}(fD_a,fD_b;\hat{e}_a,\hat{e}_b)\,,
\label{residualcorr}
\ee
where the overlap reduction functions (ORFs) are given by
\begin{align}
    \label{gammaIV_lm}
    \gamma_{\ell m}^{I,V}(fD_a,fD_b;\hat{e}_a,\hat{e}_b) 
  &= 
  \sum_{\ell_1 m_1 \ell_2 m_2}
Y_{\ell_1 m_1}(\hat{e}_a)  Y^*_{\ell_2 m_2}(\hat{e}_b) \int_{S^2} \d \hat{k}\, Y_{\ell m}(\hat{k})\, \mathbb{J}^{I,V}_{\ell_1 m_1 \ell_2 m_2} (fD_a,fD_b,\hat{k})
    \,, \\
    \label{gammaQU_lm}
    \gamma_{\ell m}^{Q\pm iU}(fD_a,fD_b;\hat{e}_a,\hat{e}_b) 
  &= 
  \sum_{\ell_1 m_1 \ell_2 m_2}
Y_{\ell_1 m_1}(\hat{e}_a)  Y^*_{\ell_2 m_2}(\hat{e}_b) \int_{S^2} \d \hat{k}\;_{\pm 4}Y_{\ell m}(\hat{k})\, \mathbb{J}^{Q\pm iU}_{\ell_1 m_1 \ell_2 m_2} (fD_a,fD_b,\hat{k})\,.
\end{align}
\ew
Eqs.~(\ref{gammaIV_lm}) and (\ref{gammaQU_lm}) are the most general ORFs for a pair of Galactic pulsars $a$ and 
$b$, respectively, at distances $D_a$ and $D_b$ from the Earth. 

\section{Calculation of $J^A_{\ell m}(fD,\hat{k})$}
\label{spectrum2}

Now we calculate the contribution of a $k$-mode to the redshift fluctuation of a pulsar, namely 
$J^A_{\ell m}(fD,\hat{k})$ in Eq.~(\ref{JAlm}).
For convenience, we first assume that $\hat{k}$ points to the direction of the polar axis or $\hat{\mathbf{z}}$-axis. In this case, 
the basis vectors~(\ref{basisvector}) become
\begin{align}
    \mathbf{e}^{+}(\hat{k}) &= \hat{\mathbf{x}}\otimes \hat{\mathbf{x}}
                               - \hat{\mathbf{y}}\otimes \hat{\mathbf{y}}\,, \nonumber \\
    \mathbf{e}^{\times}(\hat{k}) &= \hat{\mathbf{x}}\otimes \hat{\mathbf{y}}
                                    + \hat{\mathbf{y}}\otimes \hat{\mathbf{x}} \,,                         
\end{align}
and we have
\begin{align}
&\hat{e}=\sin\theta\cos\phi\, \hat{\mathbf{x}} + \sin\theta\sin\phi\, \hat{\mathbf{y}} +\cos\theta\, \hat{\mathbf{z}}\,,\\
&Y_{LM}(\hat{k})= \sqrt{\frac{2L+1}{4\pi}} \delta_{M0}\,.
\end{align}
This gives
\be
d^{ij} \mathbf{e}^{R,L}_{ij}(\hat{k})= 4\sqrt{\frac{\pi}{15}}Y_{2\pm2}(\hat{e})\,,
\ee
where the helicity $R$ takes the value of $2$ and $L$ the value of $-2$.
Hence, we obtain 
\bw
\be
\label{JAlm2}
J^{R,L}_{\ell m}(fD,\hat{\mathbf{z}})= \sum_L 4\pi\,i^{L} \sqrt{\frac{2L+1}{15}} \int_0^{2\pi fD/c}\d x\,e^{ix} j_{L}(x)
       \int_{S^2} \d \hat{e}\; Y_{2\pm2}(\hat{e})\; Y_{L0}(\hat{e})\; Y^*_{\ell m}(\hat{e})\,.
\ee
From Eq.~(\ref{eq:threeJ}), we have
\be
    \int_{S^2} \d \hat{e}\; Y_{2\pm2}(\hat{e})\; Y_{L0}(\hat{e})\; Y^*_{\ell m}(\hat{e})=
    (-1)^m\sqrt{\frac{5(2L+1)(2\ell+1)}{4\pi}}
    \begin{pmatrix}
          2 &&   L  &&  \ell \\
          0 && 0  &&  0 
    \end{pmatrix}
    \begin{pmatrix}
        2  && L  &&  \ell \\
       \pm2 && 0 &&  -m
    \end{pmatrix} \,,
\ee
which vanishes unless $m=\pm 2$ and $L=\ell-2,\ell,\ell+2$. These nonzero integral values are given by
\begin{align}
    \int_{S^2} \d \hat{e}\; Y_{2\pm2}(\hat{e})\; Y_{\ell-2,0}(\hat{e})\; Y^*_{\ell \pm2}(\hat{e})&= {1\over4}\sqrt{\frac{15}{2\pi}}
    \left[\frac{(\ell-1)\ell(\ell+1)(\ell+2)}{(2\ell-3)(2\ell-1)^2(2\ell+1)}\right]^{1\over2},\nonumber \\
    \int_{S^2} \d \hat{e}\; Y_{2\pm2}(\hat{e})\; Y_{\ell0}(\hat{e})\; Y^*_{\ell \pm2}(\hat{e})&= - {1\over2}\sqrt{\frac{15}{2\pi}}
    \left[\frac{(\ell-1)\ell(\ell+1)(\ell+2)}{(2\ell-1)^2(2\ell+3)^2}\right]^{1\over2},\nonumber \\
    \int_{S^2} \d \hat{e}\; Y_{2\pm2}(\hat{e})\; Y_{\ell+2,0}(\hat{e})\; Y^*_{\ell \pm2}(\hat{e})&= {1\over4}\sqrt{\frac{15}{2\pi}}
    \left[\frac{(\ell-1)\ell(\ell+1)(\ell+2)}{(2\ell+1)(2\ell+3)^2(2\ell+5)}\right]^{1\over2}.
\end{align}
Substituting them in Eq.~(\ref{JAlm2}), we obtain
\be
J^{R,L}_{\ell m}(fD,\hat{\mathbf{z}})=-\delta_{m\pm2} 2\pi\, i^{\ell} 
      \sqrt{\frac{(2\ell+1)}{8\pi}\frac{(\ell+2)!}{(\ell-2)!}} \int_0^{2\pi fD/c}\d x\,e^{ix} 
      \left[\frac{j_{\ell-2}(x)}{(2\ell-1)(2\ell+1)} +\frac{2j_{\ell}(x)}{(2\ell-1)(2\ell+3)}+\frac{j_{\ell+2}(x)}{(2\ell+1)(2\ell+3)}\right].
\ee
\ew
This can be cast into a compact form by using the recursion relation, $j_{\ell}(x)/x=[j_{\ell-1}(x)+j_{\ell+1}(x)]/(2\ell+1)$, which gives
\begin{align}
J^{R,L}_{\ell m}(fD,\hat{\mathbf{z}})=&-\delta_{m\pm2} 2\pi\, i^{\ell} 
      \sqrt{\frac{(2\ell+1)}{8\pi}\frac{(\ell+2)!}{(\ell-2)!}} \;\times \nonumber \\
   &   \int_0^{2\pi fD/c}\d x\,e^{ix} \frac{j_{\ell}(x)}{x^2}.
\end{align}
Through a three-dimensional rotation that takes the $\hat{\mathbf{z}}$-axis into the direction $\hat{k}$, 
we can relate~\cite{chu21}
\be
J^{R,L}_{\ell m}(fD,\hat{k}) =  \sum_{m'} D^{\ell\,*}_{m' m}(-\alpha,-\theta,-\phi) J^{R,L}_{\ell m'}(fD,\hat{\mathbf{z}})\,,
\ee
where $\hat{k}=(\theta,\phi)$. Here, the Wigner-D matrix is given by
\be
D^{\ell}_{m' m}(-\alpha,-\theta,-\phi)
    = \sqrt{\frac{4\pi}{2\ell+1}} {}_{-m'}Y_{\ell m}(\theta,\phi) e^{i m' \alpha} \,,
\ee
where $e^{i m' \alpha}$ is a redundant phase that reflects a remaining degree of freedom 
in the rotation about the $\hat{k}$-direction.
Hence, we have
\begin{align}
J^R_{\ell m}(fD,\hat{k}) &= D^{\ell\,*}_{2 m}(-\alpha,-\theta,-\phi) J^R_{\ell 2}(fD,\hat{\mathbf{z}})\,, \nonumber \\
J^L_{\ell m}(fD,\hat{k}) &= D^{\ell\,*}_{-2 m}(-\alpha,-\theta,-\phi) J^L_{\ell -2}(fD,\hat{\mathbf{z}})\,,
\label{Jresult}
\end{align}
noting that $J^R_{\ell 2}(fD,\hat{\mathbf{z}})=J^L_{\ell -2}(fD,\hat{\mathbf{z}})$ 
and $e^{i m' \alpha}$ will not appear in physical observables.

\section{Overlap Reduction Functions in the Celestial Coordinates}
\label{celestial}

Inserting the results~(\ref{Jresult}) into Eq.~(\ref{Jdefination}), we have
\bw
\begin{align}   
\mathbb{J}^{I,V}_{\ell_1 m_1 \ell_2 m_2} (fD_a,fD_b,\hat{k})&= 
(-1)^{m_1} \left[{}_{2} Y_{\ell_1 -m_1}(\hat{k}) {}_{-2} Y_{\ell_2 m_2}(\hat{k}) \pm
{}_{-2} Y_{\ell_1 -m_1}(\hat{k}) {}_{2} Y_{\ell_2 m_2}(\hat{k})\right] J_{\ell_1}(fD_a) J^*_{\ell_2}(fD_b)\,, \nonumber \\
\mathbb{J}^{Q\pm iU}_{\ell_1 m_1 \ell_2 m_2} (fD_a,fD_b,\hat{k})&= 
(-1)^{m_1} {}_{\mp 2} Y_{\ell_1 -m_1}(\hat{k}) {}_{\mp 2} Y_{\ell_2 m_2}(\hat{k})\,
e^{\pm i4\alpha} J_{\ell_1}(fD_a) J^*_{\ell_2}(fD_b)\,,
\end{align}
\ew
where we have defined the function
\be
J_{\ell}(fD)= \sqrt{2}\,\pi\, i^{\ell} \sqrt{\frac{(\ell+2)!}{(\ell-2)!}} \int_0^{2\pi fD/c}\d x\,e^{ix} \frac{j_{\ell}(x)}{x^2},
\label{JlfD}
\ee
and the phase factors $e^{\pm i4\alpha}$ are resulted from a rotation of angle $\alpha$ about the $\hat{k}$-direction on the spin-4 objects, $\mathbb{J}^{Q\pm iU}_{\ell_1 m_1 \ell_2 m_2}$, respectively. Simultaneously, the spin-4 spherical harmonics $_{\pm 4}Y_{\ell m}(\hat{k})$ in Eq.~(\ref{gammaQU_lm}) are augmented by the same phase factors of opposite signs 
$e^{\mp i4\alpha}$ under the rotation, which exactly cancel 
$e^{\pm i4\alpha}$ from $\mathbb{J}^{Q\pm iU}_{\ell_1 m_1 \ell_2 m_2}$.
Using Eq.~(\ref{eq:threeJ}) and the property~(\ref{reflection}),
the ORFs in Eqs.~(\ref{gammaIV_lm}) and (\ref{gammaQU_lm}) become
\bw
\begin{align}
    \gamma_{\ell m}^{I,V}(fD_a,fD_b;\hat{e}_a,\hat{e}_b) 
  = \sum_{\ell_1 m_1 \ell_2 m_2}  
  & (-1)^{m_1} \left[1 \pm (-1)^{\ell+\ell_1+\ell_2}\right] J_{\ell_1}(fD_a) J^*_{\ell_2}(fD_b) \;
  Y_{\ell_1 m_1}(\hat{e}_a)  Y^*_{\ell_2 m_2}(\hat{e}_b) \, \times \nonumber \\
  & \sqrt{\frac{(2\ell +1)(2\ell_1 +1)(2\ell_2 +1)}{4\pi}}
    \begin{pmatrix}
          \ell &&   \ell_1  &&  \ell_2 \\
            0  &&     -2     &&  2
    \end{pmatrix}
    \begin{pmatrix}
        \ell  && \ell_1  &&  \ell_2 \\
          m &&  -m_1  &&  m_2 
    \end{pmatrix} \,, 
    \label{gammaIV_lm_final}\\
    \gamma_{\ell m}^{Q\pm iU}(fD_a,fD_b;\hat{e}_a,\hat{e}_b) 
  = \sum_{\ell_1 m_1 \ell_2 m_2}  
  & (-1)^{m_1} J_{\ell_1}(fD_a) J^*_{\ell_2}(fD_b) \;
  Y_{\ell_1 m_1}(\hat{e}_a)  Y^*_{\ell_2 m_2}(\hat{e}_b) \, \times \nonumber \\
  & \sqrt{\frac{(2\ell +1)(2\ell_1 +1)(2\ell_2 +1)}{4\pi}}
    \begin{pmatrix}
          \ell &&   \ell_1  &&  \ell_2 \\
      \mp 4  &&  \pm 2  &&  \pm 2
    \end{pmatrix}
    \begin{pmatrix}
        \ell  && \ell_1  &&  \ell_2 \\
          m &&  -m_1  &&  m_2 
    \end{pmatrix} \,. \label{gammaQU_lm_final}
\end{align}
\ew

The two lowest moments, $\gamma_{00}^I$ and $\gamma_{00}^V$, 
select the unpolarized and the circularly polarized components of an isotropic SGWB, respectively. 
When $\ell=m=0$, the Wigner-3j symbol is proportional to 
$\delta_{\ell_1 \ell_2} \delta_{m_1 m_2}$. It immediately gives us 
\be
\gamma_{00}^V=0\,,
\ee
whereas 
\begin{align}
\gamma_{00}^I&= \sum_l \frac{2l+1}{4\pi} C_l P_l(\hat{e}_a\cdot\hat{e}_b)\quad {\rm with}\label{gammaI00_Cl}\\
C_l&\equiv \frac{1}{\sqrt{\pi}} J_l(fD_a) J^*_l(fD_b)  \,, \label{ClJJ} 
\end{align}
which depends solely on the separation angle as expected for an isotropic SGWB.
The power spectrum $C_l$ has an analytic form under the limit that $fD_a\gg c$ and $fD_b\gg c$~\cite{qin19,ng21}. Using the integral result
\be
 \int_0^\infty \d x\,e^{ix} \frac{j_l(x)}{x^2}=2 i^{l-1} \frac{(l-2)!}{(l+2)!}\,,
\ee
Eq.~(\ref{JlfD}) can be approximated as
\be
J_l(fD)\vert_{fD/c\rightarrow\infty}=2^{3/2}\pi\, i^{2l-1} \sqrt{\frac{(l-2)!}{(l+2)!}}\,,
\ee
which gives
\be
C_l=\frac{8\pi^{3/2}}{(l+2)(l+1)l(l-1)}\,.
\ee
It was shown~\cite{gair14} that the $\gamma_{00}^I$~(\ref{gammaI00_Cl}) with this $C_l$ reproduces the Hellings and Downs curve for the quadrupolar interpulsar correlations~\cite{downs}. For finite $fD_a$ and $fD_b$, one needs to calculate numerically the power spectrum~(\ref{ClJJ}) to get corrections to the Hellings and Downs curve~\cite{ng21}.

For higher multipole moments, we can easily set up a numerical scheme, similar to that in Ref.~\cite{chu2107}, to compute the ORFs in Eqs.~(\ref{gammaIV_lm_final}) and (\ref{gammaQU_lm_final}) for any pair of Galactic pulsars on the sky with known distances and coordinates, $(D_a, \theta_a, \phi_a)$ and $(D_b, \theta_b, \phi_b)$. 
The factor $J_{\ell_1}(fD_a) J^*_{\ell_2}(fD_b)$ is generally a complex number. 
When $fD_a\gg c$ and $fD_b\gg c$, we can approximate it as
\be
J_{\ell_1}(fD_a) J^*_{\ell_2}(fD_b) 
\simeq 8\pi^2 (-1)^{\ell_1+\ell_2} \sqrt{\frac{(\ell_1-2)!\,(\ell_2-2)!}{(\ell_1+2)!\,(\ell_2+2)!}}\,.
\label{Jl1Jl2}
\ee

\section{Overlap Reduction Functions in the Computational Frame}
\label{computational}

To compare the present method with previous works~\cite{anholm,mingar13,gair14,kato16,mingar14,chu2107}, we also compute the ORFs 
in the so-called computational frame: pulsar $a$ is placed along the $\hat{\mathbf{z}}$-axis
while pulsar $b$ is in the $\hat{\mathbf{x}}$-$\hat{\mathbf{z}}$ plane. Then, their polar coordinates are given by
\be
\hat{e}_a=(0,0),\quad \hat{e}_b=(\zeta,0)\,,
\ee
where $\zeta$ is their separation angle, and we have
\begin{align}
Y_{\ell_1 m_1}(\hat{e}_a)&= \sqrt{\frac{2\ell_1+1}{4\pi}} \delta_{m_1 0}, \nonumber \\
Y^*_{\ell_2 m_2}(\hat{e}_b)&=Y^*_{\ell_2 m_2}(\zeta,0)=Y_{\ell_2 m_2}(\zeta,0).
\end{align}
Hence, Eqs.~(\ref{gammaIV_lm_final}) and (\ref{gammaQU_lm_final}) simplify to
\bw
\begin{align}
    \gamma_{\ell m}^{I,V}(fD_a,fD_b,\zeta) 
  = \sum_{\ell_1 \ell_2} 
  &  (-1)^m \frac{2\ell_1+1}{4\pi} \left[1 \pm (-1)^{\ell+\ell_1+\ell_2}\right] J_{\ell_1}(fD_a) J^*_{\ell_2}(fD_b) \;
  Y_{\ell_2 m}(\zeta,0) \, \times \nonumber \\
  & \sqrt{(2\ell +1)(2\ell_2 +1)}
    \begin{pmatrix}
          \ell &&   \ell_1  &&  \ell_2 \\
            0  &&     -2     &&  2
    \end{pmatrix}
    \begin{pmatrix}
        \ell  && \ell_1  &&  \ell_2 \\
          m &&    0       &&  -m
    \end{pmatrix} \,, 
    \label{gammaIV_lm_com}\\
    \gamma_{\ell m}^{Q\pm iU}(fD_a,fD_b,\zeta) 
  = \sum_{\ell_1 \ell_2} 
  & (-1)^m \frac{2\ell_1+1}{4\pi} J_{\ell_1}(fD_a) J^*_{\ell_2}(fD_b) \;
     Y_{\ell_2 m}(\zeta,0) \, \times \nonumber \\
  & \sqrt{(2\ell +1)(2\ell_2 +1)}
    \begin{pmatrix}
          \ell &&   \ell_1  &&  \ell_2 \\
      \mp 4  &&  \pm 2  &&  \pm 2
    \end{pmatrix}
    \begin{pmatrix}
        \ell  && \ell_1  &&  \ell_2 \\
          m &&     0     &&   -m 
    \end{pmatrix} \,. \label{gammaQU_lm_com}
\end{align}
\ew
Using the property~(\ref{reflection}), it is straightforward to show that the ORFs in the computational frame have the conjugate relations:
\begin{align}
\gamma_{\ell -m}^{I}&=(-1)^{m} \gamma_{\ell m}^{I}\,,
\label{Iconjugate}\\
\gamma_{\ell -m}^{V}&=(-1)^{m+1}\gamma_{\ell m}^{V}\,,
\label{Vconjugate}\\
\gamma_{\ell -m}^{Q\pm iU}&=(-1)^{m} \gamma_{\ell m}^{Q\mp iU}\,.
\label{QUconjugate}
\end{align}
We have computed numerically the ORF multipole moments in Eqs.~(\ref{gammaIV_lm_com}) and (\ref{gammaQU_lm_com}). 
For $fD_a\gg c$ and $fD_b\gg c$, we reproduce the results for the intensity and circular-polarization ORFs in 
Refs.~\cite{anholm,mingar13,gair14,kato16}. For $fD_a=fD_b=10c$, we confirm the contribution of the pulsar term to the ORFs on small angular scales~\cite{mingar14,chu2107} 
and reproduce the results for the linear-polarization ORFs~\cite{chu2107}. 
We note that the algorithm in the present work is less complicated and more efficient than that in Ref.~\cite{chu2107}. 

\section{Bipolar spherical harmonics coefficients}
\label{biposh}

We have derived the ORFs using the harmonic-space method. Let us go back to the power spectrum in the timing-residual correlation function~(\ref{EnAv}), which can be also expanded in terms of bipolar spherical harmonics (BiPoSHs)
$\left\{Y_{\ell_1}(\hat{e}_a) \otimes Y_{\ell_2}(\hat{e}_b)\right\}_{\ell m}$ as~\cite{book:Varshalovich} 
\be
\langle r(t_a,\hat{e}_a)r(t_b,\hat{e}_b) \rangle 
=\sum_{\ell_1 \ell_2 \ell m} A^{\ell m}_{\ell_1 \ell_2} \left\{Y_{\ell_1}(\hat{e}_a) \otimes Y_{\ell_2}(\hat{e}_b)\right\}_{\ell m}\,,
\ee
where $A^{\ell m}_{\ell_1 \ell_2}$ are the expansion coefficients and
\bw
\be
\left\{Y_{\ell_1}(\hat{e}_a) \otimes Y_{\ell_2}(\hat{e}_b)\right\}_{\ell m}=
\sum_{m_1 m_2} (-1)^{m_1} \sqrt{2\ell+1} 
  \begin{pmatrix}
        \ell  && \ell_1  &&  \ell_2 \\
          m &&  -m_1  &&  m_2 
  \end{pmatrix}
  Y_{\ell_1 m_1}(\hat{e}_a)  Y^*_{\ell_2 m_2}(\hat{e}_b)\,,
\ee
which satisfy the orthogonal condition,
\be
\int_{S^2} \d \hat{e}_a\int_{S^2} \d \hat{e}_b\; 
\left\{Y_{\ell_1}(\hat{e}_a) \otimes Y_{\ell_2}(\hat{e}_b)\right\}_{\ell m}
\left\{Y_{\ell'_1}(\hat{e}_a) \otimes Y_{\ell'_2}(\hat{e}_b)\right\}^*_{\ell' m'} =
\delta_{\ell_1\ell'_1} \delta_{\ell_2\ell'_2} \delta_{\ell \ell'} \delta_{m m'}\,,
\ee
where we have used the relation~(\ref{sum3j}).
The BiPoSH coefficients is thus related to the power spectrum by
\be
 A^{\ell m}_{\ell_1 \ell_2}= \sum_{m_1 m_2}
(-1)^{m_1} \sqrt{2\ell+1} 
  \begin{pmatrix}
        \ell  && \ell_1  &&  \ell_2 \\
          m &&  -m_1  &&  m_2 
  \end{pmatrix}
  \langle a_{\ell_1 m_1} a^*_{\ell_2 m_2} \rangle\,.
\ee
From Eqs.~(\ref{EnAv}) and (\ref{residualcorr}), we have
\be
\langle a_{\ell_1 m_1} a^*_{\ell_2 m_2} \rangle =
  \intinf \frac{\d f}{(2\pi f)^2}\,(1-e^{-2\pi i f t_a}) (1-e^{2\pi i f t_b}) 
   \sum_{X=\{I,V,Q\pm iU\}}
   \sum_{\ell m}  X_{\ell m}(f) {\tilde\gamma}_{\ell m,\ell_1 m_1 \ell_2 m_2}^{X}\,,
\ee
where 
\be
\gamma_{\ell m}^{X}(fD_a,fD_b;\hat{e}_a,\hat{e}_b)=\sum_{\ell_1 m_1 \ell_2 m_2}
 {\tilde\gamma}_{\ell m,\ell_1 m_1 \ell_2 m_2}^{X} Y_{\ell_1 m_1}(\hat{e}_a)  Y^*_{\ell_2 m_2}(\hat{e}_b)\,.
\ee
Using the ORFs in Eqs.~(\ref{gammaIV_lm_final}) and (\ref{gammaQU_lm_final}), 
we can then explicitly write the BiPoSH coefficients as
\begin{align}
A^{\ell m}_{\ell_1 \ell_2} =& \intinf \frac{\d f}{(2\pi f)^2}\,(1-e^{-2\pi i f t_a}) (1-e^{2\pi i f t_b}) 
 \sqrt{\frac{(2\ell_1 +1)(2\ell_2 +1)}{4\pi}} J_{\ell_1}(fD_a) J^*_{\ell_2}(fD_b) \,\times \nonumber \\
 & \left\{
 I_{\ell m} \left[1 + (-1)^{\ell+\ell_1+\ell_2}\right] 
      \begin{pmatrix}
          \ell &&   \ell_1  &&  \ell_2 \\
            0  &&     -2     &&  2
    \end{pmatrix}
 + V_{\ell m} \left[1 - (-1)^{\ell+\ell_1+\ell_2}\right] 
     \begin{pmatrix}
          \ell &&   \ell_1  &&  \ell_2 \\
            0  &&     -2     &&  2
    \end{pmatrix}
    \right. \nonumber \\
& \left. + (Q+iU)_{\ell m}
     \begin{pmatrix}
          \ell &&   \ell_1  &&  \ell_2 \\
           - 4  &&   2       &&  2
    \end{pmatrix}
+(Q-iU)_{\ell m}
     \begin{pmatrix}
          \ell &&   \ell_1  &&  \ell_2 \\
            4  &&   -2       &&  -2
    \end{pmatrix}
\right\}\,,
\end{align}
\ew
which shows that the intensity and circular-polarization anisotropies induce even-parity and odd-parity BiPoSHs respectively, whose parity is defined by the sum, $\ell+\ell_1+\ell_2$. This selection rule has been pointed out in Refs.~\cite{kam1,kam2}, though the authors considered a limiting case with $fD_a\gg c$ and $fD_b\gg c$. 
Using the property~(\ref{reflection}), under the approximation~(\ref{Jl1Jl2}) we have 
$A^{\ell m}_{\ell_1 \ell_2}=A^{\ell m}_{\ell_2 \ell_1}$. 
The BiPoSH coefficients induced by the linear-polarization anisotropy are derived for the first time in this work.
These BiPoSH coefficients can be used to construct optimal estimators for testing the statistical isotropy of the SGWB intensity and polarization~\cite{kam1,kam2}.

\section{Conclusion}
\label{conclusion}

We have studied the pulsar-timing-array observation of the Stokes parameters of an anisotropic stochastic gravitational-wave background, based on the spherical harmonic expansion of the pulsar timing residual.
A numerical scheme to compute the overlap reduction functions (ORFs) and
the bipolar spherical harmonics (BiPoSH) coefficients has been developed.
We have used the Sachs-Wolfe line-of-sight integral for the time residual of a Galactic millisecond pulsar, which properly accounts for the power contribution of the pulsar term to  the ORF multipoles at small angular separation of the pulsar pair. Using a spherical harmonic analysis, we can compute an ORF multipole with a desired angular resolution. 
Our method also gives ORFs for a Galactic pulsar pair at different distances from the Earth. Indeed, the Sachs-Wolfe line-of-sight integral can be generalized to considering extragalactic pulsars at high redshifts.
In future pulsar-timing-array observation on a large number of pulsars, 
our method provides a fast algorithm for computing accurate ORF multipoles
and BiPoSH coefficients.

\begin{acknowledgments}
This work was supported in part by the Ministry of Science and Technology (MOST) of Taiwan, Republic of China, under
Grants No. MOST 110-2112-M-032 -007 (G.C.L.) and No. MOST 110-2112-M-001-036 (K.W.N.).
\end{acknowledgments}

\appendix
\bw
\section{Spin-Weighted Spherical Harmonics}
\label{sec:spinweight}
The explicit form of the spin-weighted spherical harmonics that we use is
\begin{align}
{}_{s}Y_{\ell m}(\theta,\phi) = 
    (-1)^{m}
    e^{im\phi}
    \sqrt{\frac{(2\ell+1)}{(4\pi)}\frac{(\ell+m)!(\ell-m)!}{(\ell+s)!(\ell-s)!}}
    \sin^{2\ell}\!\left(\frac{\theta}{2}\right)
    \sum_{r}
    \binom{\ell-s}{r} \binom{\ell+s}{r+s-m}
    (-1)^{\ell-r-s}
    \cot^{2r+s-m}\!\left(\frac{\theta}{2}\right) \,.
\end{align}
\ew
When $s=0$, it reduces to the ordinary spherical harmonics,
\be
Y_{\ell m}(\hat{n})=\sqrt{\frac{(2\ell+1)}{(4\pi)}\frac{(\ell-m)!}{(\ell+m)!}}P^m_\ell(\cos \theta) e^{i m \phi} \,.
\ee

Spin-weighted spherical harmonics satisfy the orthogonal relation,
\be
    \int_{S^2} \d{\hat{n}}\; {}_{s}Y^*_{\ell m}(\hat{n}){}_{s}Y_{\ell' m'}(\hat{n})
    = \delta_{\ell \ell'} \delta_{m m'} \,,
\ee
and the completeness relation,
\begin{align}
    \sum_{\ell m} {}_{s}Y^*_{\ell m}(\hat{n}){}_{s}Y_{\ell m}(\hat{n}')
    &= \delta(\hat{n}-\hat{n}') \nonumber \\
    &= \delta(\phi-\phi')\delta(\cos\theta-\cos\theta')\,.
\end{align}
Its complex conjugate is
\be
{}_{s}Y^*_{\ell m}(\hat{n}) =  (-1)^{s+m} {}_{-s}Y_{\ell -m}(\hat{n}) \,,
\label{Yconjugate}
\ee
and its parity is given by
\be
{}_{s}Y_{\ell m}(-\hat{n}) \equiv {}_{s}Y_{\ell m}(\pi-\theta,\phi+\pi)=(-1)^{\ell} {}_{-s}Y_{\ell m}(\hat{n}) \,.
\label{Yparity}
\ee
Also, we have the spherical wave expansion:
\be
\label{eq:swexpansion}
e^{i\vec{k}\cdot\vec{r}} = 4\pi \suml \summ i^\ell j_\ell(kr) Y_{\ell m}^*(\hat{k}) Y_{\ell m}(\hat{r}) \,,
\ee
where $j_\ell(x)$ is the spherical Bessel function. 

We can calculate the integral of a product of three spin-weighted spherical harmonics using the formula:
\bw
\be
    \int_{S^2} \d \hat{e} \;
    {}_{s_1} Y_{l_1 m_1}(\hat{e})\; {}_{s_2}\!Y_{l_2 m_2}(\hat{e})\; {}_{s_3}\!Y_{l_3 m_3}(\hat{e})=
    \sqrt{\frac{(2l_1+1)(2l_2+1)(2l_3+1)}{4\pi}}
    \begin{pmatrix}
          l_1 &&   l_2  &&  l_3 \\
        -s_1 && -s_2  &&  -s_3 
    \end{pmatrix}
    \begin{pmatrix}
        l_1  && l_2  &&  l_3 \\
       m_1 && m_2  &&  m_3 
    \end{pmatrix} \,,
\label{eq:threeJ}
\ee
\ew
which involves two Wigner-3j symbols representing the coupling coefficients between different spherical harmonics~\cite{book:Varshalovich}. The Wigner-3j symbol is zero unless it satisfies: $l_1$, $l_2$, and $l_3$ have to meet the triangular condition, i.e.~$l_1 + l_2 \ge l_3 \ge |l_1- l_2|$, while $m_1+m_2+m_3=0$; when $m_1=m_2=m_3=0$, $l_1+l_2+l_3$ is even.  The Wigner-3j symbols have the reflection property and the summation relation:
\bw
\begin{align}
 &
 \begin{pmatrix}
        l_1 && l_2  &&  l_3 \\
       s_1  && s_2  && s_3
  \end{pmatrix} 
  = (-1)^{l_1 + l_2 +  l_3}
  \begin{pmatrix}
        l_1 && l_2  &&  l_3 \\
       -s_1  && -s_2  &&  -s_3
  \end{pmatrix}
  = (-1)^{l_1 + l_2 +  l_3}
  \begin{pmatrix}
        l_1 && l_3  &&  l_2 \\
       s_1  && s_3  && s_2
  \end{pmatrix}\,, 
\label{reflection} \\
& \nonumber \\
&
(2\ell +1)\sum_{m_1 m_2}
\begin{pmatrix}
        \ell  && \ell_1  &&  \ell_2 \\
          m &&  -m_1  &&  m_2 
\end{pmatrix}
\begin{pmatrix}
        \ell'  && \ell_1  &&  \ell_2 \\
          m' &&  -m_1  &&  m_2 
 \end{pmatrix}
=  \delta_{\ell \ell'} \delta_{m m'}\,.
\label{sum3j}
\end{align}
\ew


\newcommand{\Authname}[2]{#2 #1} 
\newcommand{\etal}{{\it et al.}}
\newcommand{\LSC}{\Authname{Abbott}{B.~P.} \etal\, (LIGO Scientific Collaboration)}
\newcommand{\LVC}{\Authname{Abbott}{B.~P.} \etal\, (LIGO Scientific and Virgo Collaborations)}
\newcommand{\LVK}{\Authname{Abbott}{R.} \etal\, (LIGO Scientific, Virgo, and KAGRA Collaborations)}

\newcommand{\Title}[1]{}               

\newcommand{\arxiv}[1]{\href{http://arxiv.org/abs/#1}{{arXiv:}#1}}
\newcommand{\PRD}[3]{\href{https://doi.org/10.1103/PhysRevD.#1.#2}{{Phys. Rev. D} {\bf #1}, #2 (#3)}}
\newcommand{\PRL}[3]{\href{https://doi.org/10.1103/PhysRevLett.#1.#2}{{Phys. Rev. Lett.} {\bf #1}, #2 (#3)}}

\newcommand{\MNRAS}[4]{\href{https://doi.org/10.1093/mnras/#1.#4.#2}{{Mon. Not. R. Astron. Soc.} {\bf #1}, #2 (#3)}}
\newcommand{\CQGii}[5]{\href{https://doi.org/10.1088/0264-9381/#1/#4/#5}{{Class. Quant. Grav.} {\bf #1}, #2 (#3)}}
\newcommand{\CQG}[4]{\href{https://doi.org/10.1088/1361-6382/#4}{{Class. Quant. Grav.} {\bf #1}, #2 (#3)}}
\newcommand{\JCAP}[3]{\href{https://doi.org/10.1088/1475-7516/#3/#1/#2}{{J. Cosmol. Astropart. Phys.} #1 (#3) #2}}

\newcommand{\LRR}[4]{\href{https://doi.org/10.1007/#4}{{Liv. Rev. Rel.} {\bf #1}, #2 (#3)}}
\newcommand{\ApJ}[4]{\href{https://doi.org/#4}{{Astrophys. J.} {\bf #1}, #2 (#3)}}


\raggedright


\begin{thebibliography}{}

\bibitem{ligo}
    \LVC,
    \PRL{116}{061102}{2016}.

\bibitem{ligo2019}
    \LVC,
    \CQG{37}{055002}{2020}{ab685e}.

\bibitem{ligo2050}
    For examples, see M. A. Sedda {\it et al.}, \arxiv{1908.11375}; V. Baibhav {\it et al.}, \arxiv{1908.11390}; J. Baker {\it et al.}, \arxiv{1908.11410}.
    
\bibitem{pta2019}
For a review, see S. Burke-Spolaor {\it et al.}, Astron. Astrophys. Rev. {\bf 27}, 5 (2019).

\bibitem{EPTA}
L. Lentati {\it et al.}, Mon. Not. Roy. Astron. Soc. {\bf 453}, 2576 (2015).

\bibitem{NANO}
Z. Arzoumanian {\it et al.} (NANOGrav), Astrophys. J. {\bf 821}, 13 (2016).

\bibitem{PPTA}
R. M. Shannon {\it et al.}, Science {\bf 349}, 1522 (2015).

\bibitem{SKA}
G. Janssen {\it et al.}, {\it Advancing Astrophysics with the Square Kilometre Array}, PoS AASKA14 (2015) 037.

\bibitem{SKA2}
A. Weltman {\it et al.}, Publ. Astron. Soc. Austral. {\bf 37}, e002 (2020).

\bibitem{romano}
    For a review, see J. D. Romano, \arxiv{1909.00269}.
    
\bibitem{alexander}
\Authname{Alexander}{S.~H.~S.},
    \Authname{Peskin}{M.~E.}, and 
    \Authname{Sheikh-Jabbari}{M.~M.},
    \Title{Leptogenesis from gravity waves in models of inflation}
    \PRL{96}{081301}{2006}.
    
\bibitem{satoh}
\Authname{Satoh}{M.},
    \Authname{Kanno}{S.}, and 
    \Authname{Soda}{J.},
    \Title{Circular polarization of primordial gravitational waves in string-inspired inflationary cosmology}
    \PRD{77}{023526}{2008}.
    
\bibitem{sorbo}
\Authname{Sorbo}{L.},
    \Title{Parity violation in the Cosmic Microwave Background from a pseudoscalar inflaton}
    \JCAP{06}{003}{2011}.

\bibitem{crowder2013}
    \Authname{Crowder}{S.~G.},
    \Authname{Namba}{R.},
    \Authname{Mandic}{V.},
    \Authname{Mukohyama}{S.}, and 
    \Authname{Peloso}{M.},
    \Title{Measurement of parity violation in the early universe using gravitational-wave detectors}
     \href{https://doi.org/10.1016/j.physletb.2013.08.077}{Phys. Lett. B {\bf 726}, 66 (2013)}.
   
\bibitem{cusin}
\Authname{Cusin}{G.},
    \Authname{Durrer}{R.}, and 
    \Authname{Ferreira}{P.~G.},
    \Title{Polarization of a stochastic gravitational wave background through diffusion by massive structures}
    \PRD{99}{023534}{2020}.
    
\bibitem{bartolo}
    \Authname{Bartolo}{N.},
    \Authname{Bertacca}{D.},
    \Authname{ Matarrese}{S.},
    \Authname{Peloso}{M.},
    \Authname{Ricciardone}{A.},
    \Authname{Riotto}{A.}, and 
    \Authname{Tasinato}{G.},
    \Title{Characterizing the cosmological gravitational wave background: Anisotropies and non-Gaussianity}
    \PRD{102}{023527}{2020}.
   
\bibitem{pitrou}
\Authname{Pitrou}{C.},
    \Authname{Cusin}{G.}, and 
    \Authname{Uzan}{J.-P.},
    \Title{Unified view of anisotropies in the astrophysical gravitational-wave background}
    \PRD{101}{081301(R)}{2020}.
    
\bibitem{liu21}
J. Liu, R.-G. Cai, and Z.-K. Guo, Phys. Rev. Lett. {\bf 126}, 141303 (2021).

\bibitem{cai21}
R.-G. Cai, Z.-K. Guo, and J. Liu, arXiv:2112.10131.

\bibitem{nanograv}
NANOGrav Collaboration: Z. Arzoumanian {\it et al.}, Astrophys. J. Lett. {\bf 905}, L34 (2020).

\bibitem{PPTA21}
B. Goncharov {\it et al.},  Astrophys. J. Lett. {\bf 917}, L19 (2021).

\bibitem{EPTA21}
S. Chen {\it et al.},  Mon. Not. Roy. Astron. Soc. {\bf 508}, 4970 (2021).

\bibitem{IPTA22}
J. Antoniadis {\it et al.},  arXiv:2201.03980.

\bibitem{ligo2101}
R. Abbott {\it et al.} (LIGO Scientific, Virgo, and KAGRA Collaborations), Phys. Rev. D {\bf 104}, 022004 (2021).

\bibitem{ligo2103}
R. Abbott {\it et al.} (LIGO Scientific, Virgo, and KAGRA Collaborations), Phys. Rev. D {\bf 104}, 022005 (2021).

\bibitem{anholm}
M. Anholm, S. Ballmer, J. D. E. Creighton, L. R. Price, and X. Siemens, Phys. Rev. D {\bf 79}, 084030 (2009).

\bibitem{mingar13}
C. M. F. Mingarelli, T. Sidery, I. Mandel, and A. Vecchio, Phys. Rev. D {\bf 88}, 062005 (2013).

\bibitem{gair14}
J. Gair, J. D. Romano, S. Taylor, and C. M. F. Mingarelli, Phys. Rev. D {\bf 90}, 082001 (2014).

\bibitem{kato16}
R. Kato and J. Soda, Phys. Rev. D {\bf 93}, 062003 (2016).

\bibitem{mingar14}
C. M. F. Mingarelli and T. Sidery, Phys. Rev. D {\bf 90}, 062011 (2014).

\bibitem{qin19}
W. Qin, K. K. Boddy, M. Kamionkowski, and L. Dai, Phys. Rev. D {\bf 99}, 063002 (2019).

\bibitem{chu21}
Y.-K. Chu, G.-C. Liu, and K.-W. Ng, Phys. Rev. D {\bf 103}, 063528 (2021).

\bibitem{chu2107}
Y.-K. Chu, G.-C. Liu, and K.-W. Ng, Phys. Rev. D {\bf 104}, 124018 (2021).

\bibitem{ng21}
K.-W. Ng, arXiv:2106.12843.

\bibitem{taylor1}
S. R. Taylor and J. R. Gair, Phys. Rev. D {\bf 88}, 084001 (2013).

\bibitem{taylor2}
S. R. Taylor, R. van Haasteren, and A. Sesana, Phys. Rev. D {\bf 102}, 084039 (2020).

\bibitem{kam1}
S. C. Hotinli, M. Kamionkowski, and A. H. Jaffe, Open J. Astrophys. {\bf 2}, 1 (2019).

\bibitem{kam2}
E. Belgacem and M. Kamionkowski, Phys. Rev. D {\bf 102}, 023004 (2020).

\bibitem{smith1}
Y. Ali-Ha\"{\i}moud, T. L. Smith, and C. M. F. Mingarelli, Phys. Rev. D {\bf 102}, 122005 (2020).

\bibitem{smith2}
Y. Ali-Ha\"{\i}moud, T. L. Smith, and C. M. F. Mingarelli, Phys. Rev. D {\bf 103}, 042009 (2021).

\bibitem{book:BornAndWolf} 
    \Authname{Born}{M.} and \Authname{Wolf}{E.},
    {\em Principles of Optics}, 6th ed.
    (Pergamon Press, New York, 1980).
    
\bibitem{sachs}
R. K. Sachs and A. M. Wolfe, Astrophys. J. {\bf 147}, 73 (1967).

\bibitem{downs}
R. W. Hellings and G. S. Downs, Astrophys. J. {\bf 265}, L39 (1983).

    

\bibitem{book:Varshalovich} 
    \Authname{Varshalovich}{D.~A.}, \Authname{Moskalev}{A.~N.}, and \Authname{Khersonskii}{V.~K.},
    {\em Quantum Theory of Angular Momentum}
    (World Scientific, Singapore, 1988).
    

\end{thebibliography}
\end{document}